\documentclass[aps,prl,twocolumn,superscriptaddress]{revtex4-2}
\usepackage{graphicx}
\usepackage{dcolumn}
\usepackage{bm}
\usepackage{amssymb}

\begin{document}

\title{Stochastic switching in quadrupole-trap-based levitodynamic systems}
\author{Vadim Rybin}
\email[Contact author: ]{vadi.rybin@gmail.com}
\affiliation{PhysNano Department, ITMO University, St.~Petersburg, Russia}
\author{Semyon Rudyi}
\author{Dmitrii Shcherbinin}
\affiliation{PhysNano Department, ITMO University, St.~Petersburg, Russia}
\affiliation{Institute for Problems in Mechanical Engineering RAS, St.~Petersburg, Russia}
\date{\today}

\begin{abstract}
We have theoretically investigated the stochastic switching dynamics of a charged microparticle in a quadrupole trap near the principal parametric resonance at atmospheric pressure. We show that this behavior arises from a balance of thermal noise-activated escape and dissipative return between linear and nonlinear dynamical regimes of the particle motion. The switching follows an effective Gibbs distribution with an effective potential barrier between the regimes and effective system temperature, which are both experimentally accessible. Because the effective barrier depends on particle mass and charge, and since thermal noise reveals rather than obscures the transition, this exponential dependence could provide the foundation to precision noise-aware detection schemes. The system further offers a highly adjustable platform for studying activation dynamics in non-equilibrium parametric systems.
\end{abstract}

\maketitle


Parametric oscillators operating at the principal resonance, where the drive frequency is twice the natural frequency, exhibit a divergent sensitivity of their oscillation amplitude to small variations in system parameters~\cite{houri2025comparing}. Parametric modulation near the resonance point allows enhancing the sensitivity of inertial and environmental sensors~\cite{nitzan2015self, houri2025comparing}, and applying mechanical noise squeezing protocols~\cite{szorkovszky2011mechanical, rugar1991mechanical}. At the principal parametric resonance point, the system undergoes a dynamical transition characterized by a significant  growth in oscillation amplitude and an abrupt change in a phase portrait type. This qualitative change is exploited in threshold-trigger sensing~\cite{meesala2020bifurcation} and bifurcation topology tracking~\cite{xu2024sensitivity,  karabalin2011signal}, where the system response acts as a binary flag and yields a dramatic enhancement in detection sensitivity. Meanwhile, enhanced system sensitivity also enhances thermal noise impact, which is considered to be a limiting factor for all the cases mentioned~\cite{qiao2022analysis}.

The same noise vulnerability that limits conventional deterministic measurements can be turned into an advantage as a stochastic ``noise-aware'' detection scheme operating near parametric resonance conditions~\cite{qiao2022analysis}. The concept behind these approaches is to monitor the statistical properties of a parametric dynamical system at a bifurcation point, when noise activates switching between attractors; thus switching statistics becomes a measurable entity. In this paradigm, noise is no longer an impediment, but becomes an integral part of the sensing protocol. Such phenomena as intermittency and stochastic switching between attractors have already been applied to detect weak signals~\cite{qiao2022analysis}, giving the maximum signal-to-noise ratio at the finite noise level~\cite{venstra2013stochastic}. Importantly, these parametrically driven systems are controlled in a non-equilibrium regime where the measurement sensitivity is no longer bounded by the fluctuation–dissipation theorem, which holds true for the equilibrium linear response~\cite{gomes2025fluctuation}.

The practical implementation of these non-equilibrium sensing protocols requires physical systems that exhibit parametric resonance and allow bifurcation dynamics. The most extensively studied parametric oscillator platforms are the micro-electromechanical systems where mechanical oscillation of a micro-fabricated frame is driven by electrical radio-frequency actuation~\cite{moran2013review}. They are often manufactured using lithography, which enables miniaturization, mass production, and low power consumption. However, these advantages also become considerable limitations. Device geometry, resonator mass, and stiffness are predetermined, offering limited post-fabrication tunability. Operation beyond the parametric resonance point is also generally avoided, as it leads to amplitude deflection and complex nonlinear dynamics of the oscillator, causing its unpredictable instability or even failure~\cite{liu2024amplitude}.

An alternative platform, also described by the parametric oscillator model, is levitodynamic systems based on a charged particle confined in a quadrupole trap~\cite{gonzalez2021levitodynamics, mihalcea2023physics}. Such a levitated system offers ultimate tunability: particle mass and charge can be preselected and controllably changed during an experimental cycle, and buffer gas pressure and trapping field parameters are widely adjustable. In addition, the particle trajectory is accessible via optical imaging or interferometry~\cite{gonzalez2021levitodynamics, jin2024towards}, enabling direct system characterization. Under significant nonlinearity of the drag force, the levitated system dynamics can be stabilized beyond the principal parametric resonance point, providing access to new dynamical regimes and preventing particle ejection from the trap~\cite{rybin2023novel, rybin2022nano}. Apart from parametrically driven systems, levitodynamic platforms have proven to be a versatile tool for precision thermodynamic experiments, enabling the first direct observation of the Kramers turnover~\cite{rondin2017direct} and experimental implementation and control of non-equilibrium dynamics at the nanoscale~\cite{arita2023cooling}.

Our recent experimental study on a single charged microparticle motion in a quadrupole trap revealed a phase transition between linear and nonlinear dynamical regimes, accompanied by a well-defined and time-sustained transient state~\cite{rybin2025phase}. This transient state exists in a narrow range of the system parameters near the principal resonance point and shows a nonlinear growth of its sensitivity to small variations in the system parametric modulation depth. Based on the results obtained, we consider the opportunity to implement a ``noise-aware'' protocol based on transient dynamics within the quadrupole-trap-based levitodynamic system. In this Letter, we develop a theoretical framework that establishes functional relationship between the system parameters and the particle transient dynamics, thereby laying the foundation for future experimental verification of the noise-aware sensing concept.

\textit{Model details}. -- In accordance with our previous experimental finding~\cite{rybin2025phase}, here we consider the motion of a spherical uniform microparticle confined in a quadrupole trap in air at atmospheric pressure and room temperature. The particle size is much larger than the mean free path of the gas molecules, and the system Knudsen number is small ($\textrm{Kn}\ll1$), which justifies the hydrodynamic description of the dissipative forces. The thermal fluctuation of gas leads to a random momentum exchange between the trapped particle and the gas molecules, which requires stochastic description. This effect primarily manifests below the resonance point, where the particle oscillations are damped, and the system dynamics becomes driven by the gas thermal fluctuations.

\begin{figure}[t!]
\centering
\includegraphics[width=0.5\linewidth]{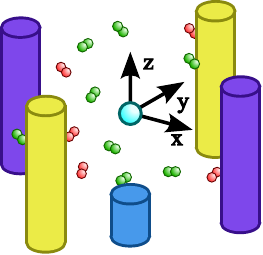}
\caption{\label{fig:trap} Schematic view of the system. A charged spherical microparticle is trapped by the quadrupole trap under normal pressure and temperature conditions. The quadrupole trap consists of four cylindrical rod electrodes, yellow and purple, powered by high alternating voltage; same colored electrodes indicate in-phase connection, differently colored electrodes are powered in antiphase. The end-cap electrode powered with a constant electric potential that compensates the particle gravity, is shown in blue. The coordinate frame is placed at the particle stable quasi-equilibrium point. The principal axis of the trap coincides with the $z$-axis; the $x$- and $y$-axes are directed toward the trap rod electrodes and form the radial plane. The red and green spheres represent air molecules, not to scale.}
\end{figure}

With these environmental conditions established, we now turn to the quadrupole trap geometry. We presume that the trap is oriented vertically along the $z$-axis and the particle gravitational force is balanced by the constant electric potential of the end-cap electrode, giving no projection on the $xy$-plane, Fig.~\ref{fig:trap}. In this case, assuming the particle motion along the $z$-axis to be negligible, we proceed to analyze the dynamics of the system in the $xy$-plane as a two-dimensional problem.

The particle dynamics is determined by the electric force from the trap potential, the viscous drag from the buffer gas, and the stochastic force due to thermal fluctuations of the gas medium. The electric potential near the trap principal axis is given by $\Phi(\textrm{x},\textrm{y},t) = V \cos(\omega t)\cdot (\textrm{x}^2 - \textrm{y}^2)/2 r_0^2$, where $V$ and $\omega$ are the amplitude and frequency of the trap electric field, and $r_0$ is the trap radius. The corresponding electric force is $\mathbf{F}^Q = -e \nabla \Phi$, with $e$ denoting the particle charge.

The viscous drag is described within the hydrodynamic approximation using the Oseen model, which for the $x$-axis is given as $\mathbf{F}^D(\dot{\textbf{x}}) = -\gamma (1 + \alpha |v|) \dot{\textbf{x}}$. Here, $\gamma = 6 \pi \mu r$ is the Stokes friction coefficient, $\mu$ is the dynamic viscosity of the buffer gas, $r$ is the particle radius, $\alpha = 3 \rho r / 8 \mu$ is the nonlinear friction coefficient, $\rho$ is the density of the buffer gas, and $|v|$ is the trapped particle instant velocity magnitude. The Oseen nonlinear correction becomes essential at the parametric resonance, where the particle velocity is sufficiently high for the Reynolds number to reach the order of unity ($\textrm{Rn}\sim1$)~\cite{le1970numerical}. 

The Brownian motion of the particle is taken into account according to the Wiener process model within discrete approximation $\mathbf{F}_i^B = \sqrt{n} \cdot \mathbf{R}_i$, where $n=2 \gamma k_B T / \Delta t$, $k_B$ is the Boltzmann constant, $T$ is the gas temperature, $\Delta t$ is the discrete time step. $\mathbf{R}_i$ are independent Gaussian random variables with zero mean and unit variance, so $\langle \mathbf{F}_i^B \rangle = 0$ and $\langle \mathbf{F}_i^B \mathbf{F}_j^B \rangle = n~\delta_{ij}$, where $\delta_{ij}$ is the Kronecker symbol, and $i, j$ are indices of the corresponding coordinates.

By introducing standard substitutions $\tau=\omega t/2$, $q=2eV/m\omega^2r_0^2$, $\beta=2\gamma/m\omega$, $\eta=\alpha\sqrt{2\omega n}/m\omega$ and normalizing the coordinate as $x=\alpha\omega \textrm{x}/2$, the equations of motion in the $xy$-plane can be written in the following form:

\begin{eqnarray}
\label{eq:eqm}
\ddot x &= -2q\cos(2\tau)x  - \beta(1+\sqrt{\dot x^2+\dot y^2})\dot x + \eta R_i,\nonumber \\
\ddot y &= +2q\cos(2\tau)y  - \beta(1+\sqrt{\dot x^2+\dot y^2})\dot y + \eta R_j.
\end{eqnarray}

As could be seen, the system dynamics is determined by three dimensionless parameters: $q$, $\beta$, and $\eta$. Here, $q$ defines the interaction between the charged microparticles and the trap quadrupole field, which is proportional to the trap voltage $V$; and it controls the depth of parametric modulation. The value of $\beta$ quantifies the rate of energy dissipation, and, while the Knudsen number is low, does not essentially depend on gas density $\rho$ (and, consequently, pressure $P$). Notably, $\beta$ can be individually tuned by selecting the particle size-to-mass ratio. The value of $\eta$ defines the degree of stochastic interaction and, in contrast, is sensitive to the residual gas density, which can serve as a variable for adjusting the intensity of Brownian motion. This separation of control parameters is particularly advantageous for practical implementation, giving an independent leverage over all the interactions in the system. One can independently adjust the noise level by changing the pressure, set the dissipation rate by choosing the right particle, and select modulation depth by choosing the trap voltage. This allows systematic exploration of the system non-equilibrium regimes within one experimental setup. 

\begin{figure*}[ht]
\centering
    \begin{minipage}{.4\linewidth}
    \includegraphics[width=0.8\linewidth]{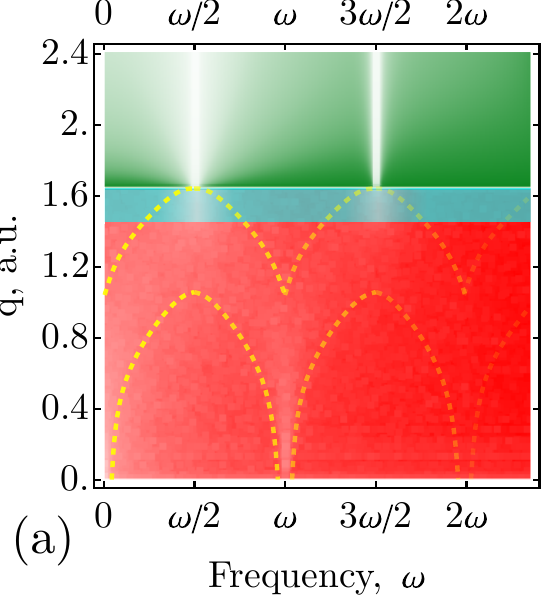}
    \end{minipage} 
    \begin{minipage}{.27\linewidth}
\includegraphics[width=1\linewidth]{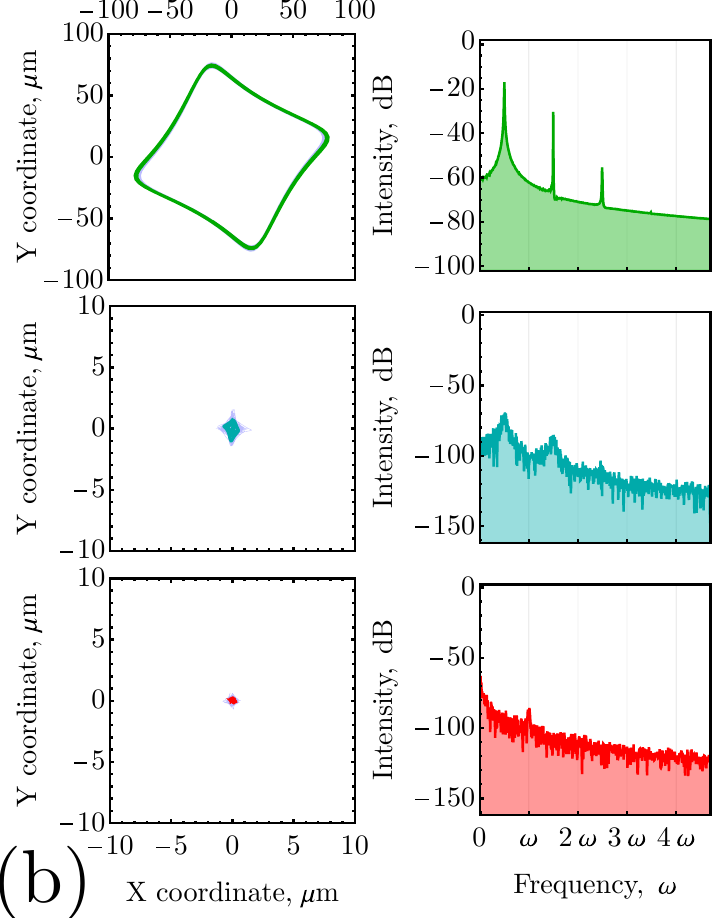}
    \end{minipage} \quad
    \begin{minipage}{.27\linewidth}
\includegraphics[width=0.93\linewidth]{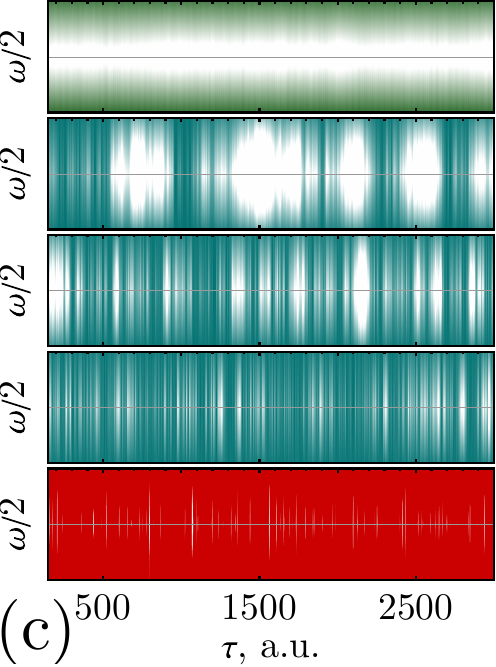}
    \end{minipage}
\caption{\label{fig:Tr-PSD-trio} (a) Spectral map in the plane of parameter $q$ and the frequency of the particle translational oscillations for $\beta=1.5$, where the color brightness encodes the intensity of spectral bands. The figure shows the transition boundary between the linear (red), stochastic switching (cyan), and nonlinear (green) regimes. Yellow dashed lines indicate regions where secular and micro-frequency motions are present. (b) The trapped particle trajectory in the radial plane and the corresponding power spectral density plot. Motion in the nonlinear regime at $q=1.659,~\beta=1.5$ is shown in green. The particle moves along a large-amplitude limit cycle, extended orbit; the spectrum is dominated by peaks at half-integer multiples of the drive frequency $(1/2+k)\omega$. Motion in the stochastic switching regime at $q=1.609,~\beta=1.5$ is shown in cyan. The particle undergoes low-amplitude oscillations near the trap center; the oscillation spectrum comprises bands at $(1/2+k)\omega$, whose intensity well exceeds the thermal noise level. Motion in linear regime at $q=0.351,~\beta=1.5$ is shown in red. The particle undergoes low-amplitude oscillations near the trap center; the spectrum is dominated by thermal noise; secular frequency $\Omega$ and micromotions $k\omega\pm\Omega$ are strongly damped. The thin blue line represents the particle trajectory, calculated over a long time interval. (c) Spectrograms of the range near the $\omega/2$, the color encodes the intensity of the spectral band, with brighter tones corresponding to higher intensity. The top green spectrogram corresponds to nonlinear regime at $q=1.659$, the middle cyan spectrograms to stochastic switching regime at $q=1.651, 1.648, 1.637$ with $T_{switch}=0.64, 0.52, 0.29$, correspondingly. The bottom red spectrogram corresponds to linear regime at $q=1.505$. The system relative stay time in the nonlinear regime, i.e., $T_{switch}$, increases with increasing $q$.}
\end{figure*}

\textit{Dynamical regimes}. -- In the case of neglected nonlinear correction of the drag force and Brownian motion, the system is reduced to a damped oscillator with parametric excitation. This oscillator is characterized by secular frequency $\Omega\propto\sqrt{q^2+\beta^2/4}$ and micromotion frequencies $k\omega\pm\Omega, k\in\mathbb{Z}$, and features a principal resonance at $2\Omega=\omega$~\cite{kovacic2018mathieu}. Depending on detuning from this condition, and due to the appearance of stochastic and nonlinear processes the system exhibits qualitatively different behavior. Far from resonance, the particle motion is damped and dominated by thermal fluctuations. At resonance, the linearized theory predicts unbounded growth of the oscillation amplitude, leading to particle ejection from the trap. However, in our case, i.e. in the presence of strong nonlinear drag, this instability is suppressed~\cite{kovacic2018mathieu, rybin2025phase}. In this scenario the dissipation rate grows nonlinearly with the particle velocity, allowing a balance between the parametric pumping and energy dissipation. The system thus enters a steady-state high-amplitude oscillatory regime with the doubled period, known as the ``extended orbit''~\cite{rybin2025phase}. 

To analyze the complete nonlinear stochastic dynamics, now we turn to the numerical solution of Eq.~\ref{eq:eqm}. The equations are solved using the It$\hat{\textrm{o}}$ method~\cite{allen2007modeling} with initial position and velocity sampled from the thermal equilibrium distributions, corresponding to the linear regime of the trapped particle motion. The total simulation time was chosen to be sufficiently long to ensure statistical convergence of the solutions and their independence of initial conditions. For the calculation, we set $\beta=1.5$ and $\eta=10^{-5}$, corresponding to the trapping of a $30$-$\mu$m polystyrene sphere at a trap field frequency of $70$~Hz and standard temperature and pressure conditions. The parameters were chosen close to those from our previous experimental work~\cite{rybin2025phase}. We first analyze the trajectory and spectral composition of the particle motion taking~$q$ as a control parameter. The results are summarized in the spectrogram shown in Fig.~\ref{fig:Tr-PSD-trio}(a). Here, the ordinate axis represents control parameter~$q$, the abscissa axis is the particle oscillation frequency normalized by drive frequency~$\omega$, and the tones encode the spectral intensity. 

\begin{figure*}[ht]
    \begin{minipage}{.3\linewidth}
    \includegraphics[width=0.95\linewidth]{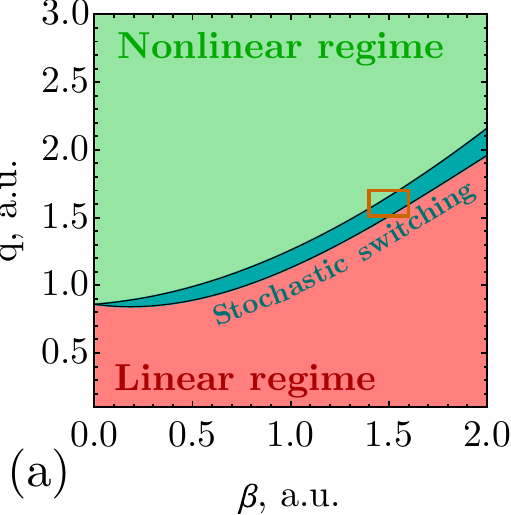}
    \end{minipage} \quad
    \begin{minipage}{.3\linewidth}
\includegraphics[width=1\linewidth]{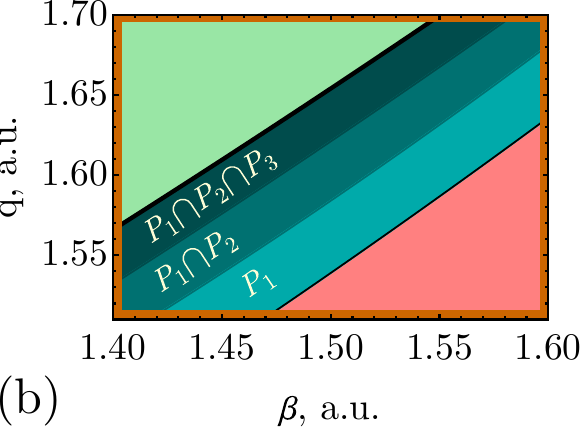}
    \end{minipage} \quad
    \begin{minipage}{.3\linewidth}
\includegraphics[width=0.95\linewidth]{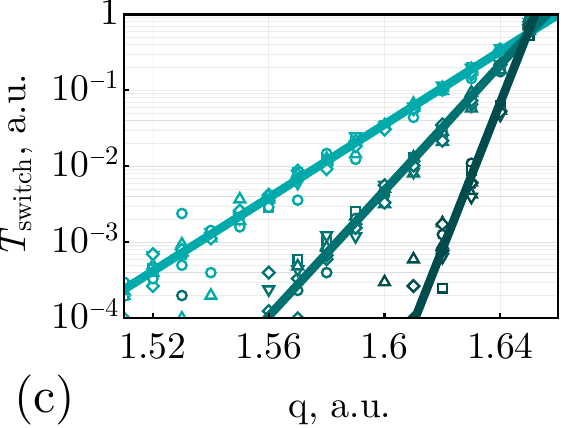}
    \end{minipage}
\caption{\label{fig:qb-diagramm} (a) Diagram of the system dynamic regimes in the $q$ and $\beta$ parameter plane; with linear, stochastic switching, and nonlinear regimes corresponding to red, cyan, and green, respectively. The boundary between the linear and stochastic switching regimes is set in terms of the emergence of the $\omega/2$ band in the particle oscillation spectrum, $T_{switch} \ge 0.001$. The boundary between the stochastic switching and nonlinear regimes is determined by the manifestation of the $\omega/2$ band in the particle oscillation spectrum and the formation of the extended orbit trajectory, $T_{switch}=1$. (b) The diagram shows a enlarged fragment highlighted by an orange rectangle for different gas pressure values, $P_1=1~\textrm{atm}, P_2=0.75~\textrm{atm}, P_3=0.5~\textrm{atm}$. A decrease in pressure narrows the stochastic switching region: $P_1$ corresponds to the lower, middle, and upper cyan zones, $P_2$ to the upper zones, and $P_3$ to the top zone. (c) The dependence of the stochastic switching constant $T_{switch}$ on the parameter $q$ on a logarithmic scale for different values of residual gas pressure. A decrease in pressure leads to narrowing the width of the stochastic switching region and to an increase in the slope of the $T_{switch}(q)$ dependence.}
\end{figure*}

At low $q$, the spectrogram reveals residual bands at secular frequency $\Omega$ and at micromotion sidebands $k\omega\pm\Omega$. These bands are strongly damped and broadened under the considered conditions. We highlight them using yellow dashed lines in the figure. We refer to this behavior as the \textit{linear regime}, with a typical trajectory and motional spectra shown in red in Fig.~\ref{fig:Tr-PSD-trio}(b). As $q$ increases, the secular and micromotion bands converge near $\omega/2$ and reinforce each other, forming the second, \textit{stochastic switching regime} (whose name is explained below). Motional spectra acquire subtle bands on $(1/2+k)\omega, k\in\mathbb{Z}$, while the oscillation amplitude remains low, cyan in Fig.~\ref{fig:Tr-PSD-trio}(b). Upon reaching a critical value of the modulation depth, $q_{crit}$, the system abruptly transits to a \textit{nonlinear regime} with the formation of an extended orbit, green in Fig.~\ref{fig:Tr-PSD-trio}(b).

In real systems, linear and stochastic switching regimes are typically characterized by strong peaks at trap field frequencies~$k\omega$~\cite{rybin2025phase}. This is caused by various types of inhomogeneities in the system, such as the asymmetry in the trap potential, inhomogeneity in the end-cap electrode field, contribution of edge effects, uncompensated tilt of the trap axis, which leads to the appearance of the particle gravitational force projection in the radial plane of the trap, and others. Nevertheless, for the purposes of this letter, we have intentionally excluded all the inhomogeneities from the equations of motion.

The key distinction between the regimes lies in the behavior of the spectral component at $\omega/2$, Fig.~\ref{fig:Tr-PSD-trio}(c). In the linear regime, the periodogram of the region near $\omega/2$ shows only thermal noise without any distinct signal. In the nonlinear regime, in turn, the $\omega/2$ band is intense and stable in time, reflecting the steady-state high-amplitude motion along the extended orbit. In the stochastic switching regime, the $\omega/2$ intensity fluctuates between high and low values, indicating stochastic switching between the linear and nonlinear attractors. Moreover, as $q$ increases within the stochastic switching region, the fraction of time spent in the high-intensity state grows monotonically, as seen from Fig.~\ref{fig:Tr-PSD-trio}(c). This behavior results from momentum transfers to the particle through random collisions with the surrounding gas molecules. When the accumulated energy is sufficient to drive the particle across the bifurcation point, the system escapes from the linear into the nonlinear regime, where the excess energy is dissipated by the viscous drag or carried back by a subsequent collision. The observed stochastic switching thus reflects a dynamical balance between escape and return, whose relative durations are controlled by detuning from the parametric resonance. This behavior provides a direct quantitative basis for characterizing the system non-equilibrium dynamics.

\textit{Stochastic switching sensitivity}. -- To quantify the stochastic switching behavior, we introduce the stochastic switching constant as $T_{switch}=T_{n}/(T_{l}+T_{n})$, where $T_{n}$ is the time the system spends in the nonlinear regime (high-intensity of $\omega/2$ band) and $T_{l}$ is the time spent in the linear regime (low-intensity of $\omega/2$ band) over a finite time interval. Thus, the ratio approaches unity, $T_{switch}\rightarrow1$, in the case of a purely nonlinear regime, and zero, $T_{switch}\rightarrow0$, in the case of a purely linear regime. Here, we define the cut-off value of a lower bound as $10^{-3}$ to ensure reliable $T_{switch}$ detection during the numerical simulation.

Taking into account the introduced stochastic switching constant, we construct a map of dynamic modes in the $q–\beta$ parameter plane; Fig.~\ref{fig:qb-diagramm}(a), where green, red, and cyan correspond to nonlinear, linear and stochastic switching regimes, respectively. The figure generalizes the results to a broad range of system parameters. As $\beta$ approaches zero, the width of the stochastic switching region also tends toward zero, which makes physical sense, as both the dissipation rate and thermal noise intensity are directly proportional to the same buffer gas parameters.

Since the stochastic switching regime is driven by thermal fluctuations, its width in the $q–\beta$ plane is controlled by noise parameter $\eta$, which can be independently set by the buffer gas density. In our case, gas density could be reduced up to an order of magnitude while keeping the Knudsen number low. The connection between the width of the stochastic switching region and gas pressures $P_1=1~\textrm{atm}, P_2=0.75~\textrm{atm}$ and $P_3=0.5~\textrm{atm}$ is illustrated in Fig.~\ref{fig:qb-diagramm}(b). As the pressure decreases, the stochastic switching region narrows, reflecting the reduction of the noise level. The quantitative dependence of stochastic switching constant, $T_{switch}$, on $q$ parameter is shown in Fig.~\ref{fig:qb-diagramm}(c) for various gas pressures. On a logarithmic scale, $T_{switch}$ exhibits a linear dependence on $q$, revealing an exponential law, $T_{switch}\propto\exp[-(q_{crit}-q)/\eta^2]$. This relationship is an effective analog of the Gibbs distribution, where the height of the effective potential barrier between the regimes takes the form $\Delta U_{eff}(q)=q_{crit}-q$, and the effective temperature of the system is $T_{eff}=\eta^2/k_B$. Both quantities are experimentally accessible from the measured dependence of $T_{switch}$ on $q$ at the given trapping and environmental parameters. 

Notably, because the height of the effective barrier depends on the particle mass and charge, the exponential sensitivity of  $T_{switch}(q)$ becomes a powerful tool for detecting small changes in these parameters. Unlike conventional measurements that rely on the amplitude response or spectral band shift, which are both limited by thermal noise, the present approach extracts information from the statistical properties of the noise-activated switching dynamics. In this sense, noise does not obscure the signal; it reveals the fine structure of the effective barrier by probing the system response across the parametric resonance point. This makes the measurement inherently robust to the noise level itself, as the sensitivity is encoded in the functional form of $T_{switch}(q)$, not in the particle coordinate itself. The price paid for this divergent sensitivity is a reduced dynamic range; the width of the stochastic switching region shrinks as the pressure decreases. Thus, the operating buffer gas density determines the trade-off between sensitivity and dynamic range.

\textit{Conclusions}. -- In this Letter, we have developed a theoretical framework that explains the phase transition between linear and nonlinear dynamical regimes recently observed experimentally in a levitated microparticle dynamics near the parametric resonance point~\cite{rybin2025phase}. The  model proposed reveals the stochastic switching dynamics that arises from a dynamical balance between noise-induced escape and dissipative return between linear and nonlinear regimes, mediated by random collisions with gas molecules. We have shown the exponential dependence of the switching statistics on the control parameter, which follows an effective analog of the Gibbs distribution for the two-state switching dynamics. The effective barrier height and effective temperature provide a direct qualitative explanation for the growing sensitivity observed in the experiment as the system approaches the critical modulation depth.

We highlight the functional relation between the particle mass and charge and the effective barrier height, which encodes information about these parameters into the statistical properties of the switching dynamics. This enhances the system sensitivity at a given dissipation and noise level and makes the stochastic switching regime promising for noise-aware detection protocols. Thus, the accuracy of measurement is derived from the functional form of the switching probability rather than from the amplitude response or the frequency shift.

Beyond possible sensing applications, the system provides a highly controllable and adjustable experimental platform for studying stochastic processes in non-equilibrium systems. It enables quantitative tests of activation dynamics, and could complement the Kramers escape theory for the case of phase transitions in systems with parametric excitation. Thus, this Letter establishes both a practical route to noise-aware sensing and a versatile testbed for fundamental studies of non-equilibrium thermodynamics.


\begin{acknowledgments}
This study was supported by Russian Science Foundation, project No. 25-22-00694.
\end{acknowledgments}


%


\end{document}